\newcommand{\nn}{\nonumber}
\newcommand{\mcA}{\mathcal{A}}

\documentclass[12pt, nofootinbib]{article}

\renewcommand{\title}[1]{\vbox{\center\LARGE{#1}}\vspace{5mm}}
\renewcommand{\author}[1]{\vbox{\center#1}\vspace{5mm}}
\newcommand{\address}[1]{\vbox{\center\em#1}}

\renewcommand{\date}[1]{\vbox{\center#1}}

\usepackage{amssymb, diffcoeff}
\usepackage{amsmath,bm}
\usepackage{amssymb}
\usepackage{graphicx}
\usepackage{amsfonts}         
\usepackage{fancybox}   
\usepackage{float}

\usepackage{fullpage}
\usepackage{lastpage}
\usepackage{enumitem}
\usepackage{wrapfig}
\usepackage{setspace}
\usepackage{graphicx}
\usepackage{calc}
\usepackage{multicol}
\usepackage{amsmath}
\usepackage[nottoc]{tocbibind}
\usepackage{subcaption}

\usepackage{enumitem}

\usepackage{slashed}

\usepackage[font=small,labelfont=bf]{caption}
\DeclareCaptionFont{tiny}{\tiny}
\usepackage{epstopdf}
\usepackage[usenames,dvipsnames]{xcolor}
\usepackage[pdftex, bookmarks={false}, pdfauthor={Diptarka Das, Santanu Mandal, Anurag Sarkar}, pdftitle={classpinstring}]{hyperref}
\hypersetup{colorlinks=true, linkcolor=BrickRed, citecolor=Violet, filecolor=OliveGreen, urlcolor=RoyalBlue, filebordercolor={.8 .8 1}, urlbordercolor={.8 .8 0}}
\usepackage{soul}
\setstcolor{Red}

\usepackage{multirow,booktabs}
\usepackage{wrapfig}

\definecolor{darkgreen}{rgb}{0,0.4,0}
\definecolor{darkred}{rgb}{0.4,0,0}
\definecolor{darkblue}{rgb}{0,0,0.4}
\definecolor{lightblue}{rgb}{.6,.6,0.9}
\definecolor{lazuli}{rgb}{0.0, 0.06, 0.54}

\definecolor{uglybrown}{rgb}{0.8,  0.7,  0.5}

\definecolor{palatinatepurple}{rgb}{0.41, 0.16, 0.38}
\definecolor{celebrationcolor}{rgb}{0.75,  0.0,  0.9}

\usepackage{mdframed}

\usepackage{framed}
\definecolor{shadecolor}{rgb}{0.90,0.90,0.90}

\usepackage{wasysym}

\numberwithin{equation}{section}

\renewcommand{\theequation}{\arabic{section}.\arabic{equation}}

\input epsf
\newcommand{\vev}[1]{\langle #1 \rangle}

\newlength{\extraspace}
\newlength{\extraspaces}
\newcommand{\td}{\text{d}}

\def\be{\begin{equation}}
\def\ee{\end{equation}}

\newcommand{\bea}{\begin{eqnarray}}
\newcommand{\eea}{\end{eqnarray}}

\def\ket#1{|#1\rangle}

\def\vev#1{\langle{#1}\rangle}

\def\CX{{\cal X}}

\def\II{\relax{I\kern-.10em I}}

\def\IB{\relax{\rm I\kern-.18em B}}

\def\ID{\relax{\rm I\kern-.18em D}}
\def\IE{\relax{\rm I\kern-.18em E}}
\def\IF{\relax{\rm I\kern-.18em F}}
\def\IG{\relax\hbox{$\inbar\kern-.3em{\rm G}$}}
\def\IGa{\relax\hbox{${\rm I}\kern-.18em\Gamma$}}
\def\IH{\relax{\rm I\kern-.18em H}}
\def\II{\relax{\rm I\kern-.18em I}}
\def\IK{\relax{\rm I\kern-.18em K}}

\def\inbar{\,\vrule height1.5ex width.4pt depth0pt}

\def\lp10{\ell_p^{10}}
\def\lp11{\ell_p^{11}}
\def\R11{R_{11}}

\def\frac#1#2{{#1 \over #2}}

\newdimen\tableauside\tableauside=1.0ex
\newdimen\tableaurule\tableaurule=0.4pt
\newdimen\tableaustep
\def\phantomhrule#1{\hbox{\vbox to0pt{\hrule height\tableaurule width#1\vss}}}
\def\phantomvrule#1{\vbox{\hbox to0pt{\vrule width\tableaurule height#1\hss}}}
\def\sqr{\vbox{%
\phantomhrule\tableaustep
\hbox{\phantomvrule\tableaustep\kern\tableaustep\phantomvrule\tableaustep}%
\hbox{\vbox{\phantomhrule\tableauside}\kern-\tableaurule}}}
\def\squares#1{\hbox{\count0=#1\noindent\loop\sqr
\advance\count0 by-1 \ifnum\count0>0\repeat}}
\def\tableau#1{\vcenter{\offinterlineskip
\tableaustep=\tableauside\advance\tableaustep by-\tableaurule
\kern\normallineskip\hbox
{\kern\normallineskip\vbox
	{\gettableau#1 0 }%
	\kern\normallineskip\kern\tableaurule}%
\kern\normallineskip\kern\tableaurule}}
\def\gettableau#1 {\ifnum#1=0\let\next=\null\else
\squares{#1}\let\next=\gettableau\fi\next}

\def\eqnn#1{\xdef #1{(\secsym\the\meqno)}\writedef{#1\leftbracket#1}%
\global\advance\meqno by1\wrlabeL#1}
\def\eqna#1{\xdef #1##1{\hbox{$(\secsym\the\meqno##1)$}}
\writedef{#1\numbersign1\leftbracket#1{\numbersign1}}%
\global\advance\meqno by1\wrlabeL{#1$\{\}$}}
\def\eqn#1#2{\xdef #1{(\secsym\the\meqno)}\writedef{#1\leftbracket#1}%
\global\advance\meqno by1$$#2\eqno#1\eqlabeL#1$$}

\global\newcount\itemno \global\itemno=0

\def\itemaut#1{\global\advance\itemno by1\noindent\item{\the\itemno.}#1}

\def\({\left(}
\def\){\right)}

\def\XX{{\bf X}}
\def\ZZ{{\bf Z}}

\def\lsim{\mathrel{\mathstrut\smash{\ooalign{\raise2.5pt\hbox{$<$}\cr\lower2.5pt\hbox{$\sim$}}}}}
\def\gsim{\mathrel{\mathstrut\smash{\ooalign{\raise2.5pt\hbox{$>$}\cr\lower2.5pt\hbox{$\sim$}}}}}

\def\overleftrightarrow#1{\vbox{\ialign{##\crcr
	$\leftrightarrow$\crcr\noalign{\kern-0pt\nointerlineskip}
	$\hfil\displaystyle{#1}\hfil$\crcr}}}

\def\overleftarrow#1{\vbox{\ialign{##\crcr
	$\leftarrow$\crcr\noalign{\kern-0pt\nointerlineskip}
	$\hfil\displaystyle{#1}\hfil$\crcr}}}

\usepackage[yyyymmdd,hhmmss]{datetime}
\newif{\ifeq}           
\eqtrue                 
\usepackage{wasysym}
\usepackage{subcaption}
\usepackage{overpic}

\begin{document}

\def\sZ{\ZZ}
\def\sX{\XX}
\def\CX{\mathfrak{X}}

\begin{titlepage}

\title{Classical string profile for a class of DDF amplitudes \\
%
}

\author{Diptarka Das, Santanu Mandal and Anurag Sarkar}

\address{ 
	{\it 
		Department of Physics, 
		Indian Institute of Technology - Kanpur, \\
		Kanpur 208016, India\\
}}

\begin{abstract}
In the critical bosonic string theory, we explicitly evaluate the three point scattering amplitude at tree level, of a photon with two massive higher spins. The massive excitations belong to states of the form $A_{-r_1}^{s_1} A_{-r_2}^{s_2}$ where $A_{-n}$ is a DDF creation operator. Next, we take the infinite ``spin'' limit to arrive at the classical string dynamics. We find a rotating ``floppy'' string lying mostly on a plane which develops a transverse kink.
\end{abstract}

\vfill
\today

\end{titlepage}
\vfill\eject
\setcounter{tocdepth}{2}    
\parskip = 0.02in

\tableofcontents

\parskip = 0.1in

\vfill\eject



\section{Introduction}

Recently there has been a resurgence of interest in amplitudes of excited string states. This is partially driven to explore certain black hole S-matrix features : realization of~{\em chaotic scattering} and emergence of~{\em thermal scales} from coarse-graining of these amplitudes \cite{Rosenhaus:2020tmv, Gross:2021gsj, Rosenhaus:2021xhm, Fukushima:2022lsd, Hashimoto:2022bll, Bianchi:2022mhs, Firrotta:2023wem, Das:2023xge, Savic:2024ock, Firrotta:2024fvi, Firrotta:2024qel, Biswas:2024unn, Biswas:2024mdu, Bhattacharya:2024szw, Bianchi:2024fsi} \footnote{Many of these studies are carried out in the DDF \cite{DelGiudice:1971yjh} basis which contains superposition of various spins. When we fix the spin there maybe no chaos due to spin universality \cite{pesando2}.}. Furthermore, it turns out that classical black hole observables can be reproduced in a post-Newtonian expansion from quantum amplitudes of massive  spin $s$ particles with the graviton of four momenta $p_3 = \hbar \bar{p}_3$. For the Kerr black hole the linearized stress energy tensor to all orders in the spin of the black hole is obtained when $s\rightarrow \infty$ as $\hbar \rightarrow 0$ \cite{Vines:2017hyw, Vaidya:2014kza,Arkani-Hamed:2017jhn,Cachazo:2017jef,Guevara:2017csg,Guevara:2018wpp,Chung:2018kqs,Johansson:2019dnu,Maybee:2019jus,Arkani-Hamed:2019ymq,Damgaard:2019lfh,Aoude:2020onz,Guevara:2020xjx,Chiodaroli:2021eug,Cangemi:2022bew,Cangemi:2023ysz,Cangemi:2023bpe,Gambino:2024uge,Bjerrum-Bohr:2025lpw}. In terms of the classical spin vector $a^\mu$ one gets the relationship: 
\begin{align}
\lim\limits_{s\rightarrow\infty, \hbar\rightarrow 0} {\cal A}\left(\Phi_1^{s} \Phi_2^{s} h_3 \right) &= h_{\mu\nu}(p_3) T^{\mu\nu}_{\text{Kerr}}(-p_3) = \left( \zeta\cdot p \right)^2 \exp \left( p_3 \cdot a \right), 
\end{align}
where the black hole mass $m^2 = p^2$. The gauge theory counter-part which is the square root of the above is called the $\sqrt{\text{Kerr}}$ amplitude. This solution corresponds to the gauge configuration resulting from a spinning charged disk \cite{Arkani-Hamed:2019ymq}. Till date, there is no derivation of the Kerr as well as $\sqrt{\text{Kerr}}$ amplitudes starting from a higher spin action. Works \cite{Cangemi:2022abk, Alessio:2025nzd} are an attempt to fill this gap starting from massive string spinning amplitudes. There are however an infinite class of states, the simplest being those lying on the leading Regge trajectory. However these produce different stress-tensors. In the open string context for instance, the classical limits of these amplitudes correspond to a rigid rotating string with a charge at the endpoint. 

To move away from this configuration one may consider other physical states. DDF \cite{DelGiudice:1971yjh} provides a way to realize a generic physical state in the string spectrum. The very recent resurgence in study of scattering amplitudes involving these states have started since the works \cite{Bianchi:2019ywd, Aldi:2021zhh}.  The DDF vertex operator that takes the form 
\begin{align}
\Phi &= \prod_r \left(\lambda_r \cdot A_{-r}\right)^{n_r} :e^{i \tilde{p}\cdot  X}: \label{eq:ddf}
\end{align}
where $\sum_r r \, n_r$ is an integer $N$. Here $A_{-r}^\mu$ are the DDF operators which follow the same algebra amongst themselves as the string oscillators, $a^\mu_{-r}$ and also commute with the Virasoro generators. This latter property guarantees physicality of these states. The mass of the state  is set by the level of the integer $N$ that is partitioned. The algebra between $A^\mu_{-r}$ and $a^\mu_{-r}$ is complicated which results in an ambiguity of assignment of spacetime spin to a generic DDF state. In general a DDF state corresponds to a superposition of massive spinning states in the target spacetime. The exact relationship at a generic level is a work in progress \cite{pesando1}. Note that in the definition \eqref{eq:ddf} the final spins are decided by the choice of $\lambda_r$. In our analysis we shall choose all of them to be same as well as $\lambda^2=0$. This excludes scalars. Furthermore, if we define $s = \sum_r n_r$, the total length of partitions, then the minimal spin in the spacetime fields is bounded from below by $s$ \cite{Bianchi:2023uby}. Therefore taking $s \rightarrow \infty$ takes the {\em spins} of spacetime fields to infinity : the classical limit. We will use this limit to extract the corresponding classical current in context of open bosonic strings. 

Recently, \cite{Alessio:2025nzd} have shown that for {\em zero-depth} states of \cite{Markou:2023ffh,Basile:2024uxn}, there is an equivalence of DDF amplitudes with the oscillator ones. These therefore correspond to amplitudes of definite spin which we also study. The DDF states that we consider are of the form: 
\begin{align}
\ket{\Phi^s} &= \left( \lambda \cdot A_{-r_1} \right)^{\alpha \, s} \left( \lambda \cdot A_{-r_2} \right)^{(1-\alpha) \, s }  :e^{i \tilde{p}\cdot  X}: \ket{0} \label{eq:ddf2}
\end{align}
The choice $r_1 = 1, \, r_2 = 2$ corresponds to zero-depth. The variable $\alpha$ parametrizes the distance from the leading Regge trajectory. When $\alpha= 1$ this is a massive state in the leading Regge trajectory. For these states the classical string profile is that of a rigid rotating string with a charge at the endpoint \cite{Cangemi:2022abk}. The goal is to see whether moving away from $\alpha=1$ may result in a configuration that approaches that of a uniformly charged rotating disk. 

In our analysis while we do not find the charged rotating disk, we do find rotating classical string configurations with more modes excited. The trajectory of the charged particle starts exploring more regions inside of the rigid rotating circle. One of the hopes could be, that with a more generic DDF state, the charge explores all of the inside of the circle in a manner which produces the same multipole expansion coefficients as that of $\sqrt{\text{Kerr}}$. Our solution is numerical, and while the string is mostly restricted to lie on a plane, there is a small region where it develops a kink in a third spatial direction. 

In section \S\ref{sec:2} we compute the tree level amplitude of a photon with two of the states as in eq.\eqref{eq:ddf2} and take the infinite spin limit. The final answer in the classical limit is expressed as a sum over products of modified Bessel functions, eq. \eqref{eq:amp}. In \S\ref{sec:3} we find the corresponding classical string profile using numerics and plot the resulting trajectories before briefly concluding in \S\ref{sec:4}. Appendix \ref{app:1} details the contractions while working out some simple examples and appendix \ref{app:15} proves some summation identities. In appendix \ref{app:16} the Regge state configuration is reviewed in our notations, and in \ref{app:2} we present further numerical details of the string profile. 

\noindent{\bf Note:} While this work was being completed, the paper \cite{ Alessio:2025nzd} appeared, which has some partial overlap with our results.

\section{String three point amplitude}\label{sec:2}

\noindent We express the DDF state of eq.\eqref{eq:ddf2} as:
\begin{align}
	\left( \lambda \cdot A_{-r_1} \right)^{N_1} \left( \lambda \cdot A_{-r_2} \right)^{N_2}  \left. |  0; \tilde{p}  \right\rangle,
\end{align}
where, $N_1 = \alpha s, N_2 = (1-\alpha) s, N_1 + N_2 = s$. The goal of this section is to analyze the $\vev{  \Phi^s_1 \Phi^s_2 \gamma  }$ amplitude in the large $s$ limit. The explicit form of the DDF creation operators involve Sch\"ur polynomials, we simplify notations as:
\begin{align} \label{HES-states}
	\Phi^s_1 &\equiv :\bigg( \sum_{m=1}^{r_1}  \frac{-i\zeta_1 \cdot \partial^{m} X_1}{(m-1)!} S_{r_1-m} \big( - \frac{i r_1}{s!} q_1 \cdot \partial^{s} X_1 \big) \bigg)^{N_1} \bigg( \sum_{n=1}^{r_2}  \frac{-i\zeta_1 \cdot \partial^{m} X_1}{(n-1)!} S_{r_2-n} \big( - \frac{i r_2}{s!} q_1 \cdot \partial^{s} X_1 \big) \bigg)^{N_2} e^{i p_1 \cdot X_1}:  \nn \\
	&=  :\big( \sum_{m=1}^{r_1} a_{1}(r_1,m) b_1(r_1,m) \big)^{N_1} \big( \sum_{n=1}^{r_2} a_{1}(r_2,n) b_1(r_2,n) \big)^{N_2} c_1: ,\nonumber \\ 
	\Phi^s_2  &\equiv :\big( \sum_{m=1}^{r_1} a_{2}(r_1,m) b_2(r_1,m) \big)^{N_1} \big( \sum_{n=1}^{r_2} a_{2}(r_2,n) b_2(r_2,n) \big)^{N_2} c_2:,  \\
\gamma &\equiv  :a_3(1,1) c_3:. \nn
\end{align}
After the contractions, the amplitude is evaluated as: 
\begin{align} \label{amphhp}
	\mathcal{A} =  \int &\text{d}z_1 ~\text{d}z_2~\text{d}z_3~\langle :\left( \sum_{m_1=1}^{r_1} a_{1}(r_1,m_1) b_1(r_1,m_1) \right)^{N_1} \left( \sum_{n_1=1}^{r_2} a_{1}(r_2,n_1) b_1(r_2,n_1) \right)^{N_2} c_1:  \nonumber \\ 
	&: \left( \sum_{m_2=1}^{r_1} a_{2}(r_1,m_2) b_2(r_1,m_2) \right)^{N_1} \left( \sum_{n_2=1}^{r_2} a_{2}(r_2,n_2) b_2(r_2,n_2) \right)^{N_2} c_2: :a_3(1,1) c_3: \rangle \nn \\
= \sum_{k}^{N_1} \sum_{l}^{N_2} & k! l! r_1^k r_2^l \binom{N_1}{k}^2 \binom{N_2}{l}^2 \left(-\zeta _1\cdot \zeta _2\right)^{k+l} \left(-\zeta _1\cdot p_3 \zeta _2\cdot p_3\right)^{-k-l+N_1+N_2-1}\nn \\
		& \left(\zeta _2\cdot f_3\cdot \zeta _1 \left(r_1 \left(N_1-k\right)+r_2 \left(N_2-l\right)\right) + \zeta _3\cdot p_1 \left( \zeta _1\cdot p_3 \zeta _2\cdot p_3 \right) \right) .
\end{align}
The details about the contractions leading to the derivation of the above r.h.s is given in the appendix \ref{app:1}. 

\subsection{Spin coefficients at finite $s$}
\noindent Following replacements \cite{Cangemi:2022abk} are made to get the spin coefficients in classical limit.
\begin{align}
		&\zeta _1\cdot \zeta _2 = \vev{k \cdot a_{(1/2)}}^2 - 1, \left(\zeta _1\cdot p_3 \zeta _2\cdot p_3\right) =  -2 \vev{k \cdot a_{(1/2)}}^2 m^2,  \, \vev{k \cdot a_{(1/2)}}^{n} = \frac{(2 s-n)!}{(2 s)!} \vev{(k \cdot a_{(s)})^{n}}, \\
		&\zeta_2 \cdot f_3 \cdot \zeta_1 =  -2 \left(\zeta _3\cdot p_1\right) \left( \vev{k \cdot a_{(1/2)}}^2 + \vev{k \cdot a_{(1/2)}} \right),\nn
\end{align}
to obtain,
\begin{align}
	\mathcal{A} = - \left(\zeta _3\cdot p_1\right) \sum_{n=0}^{2s} c_n \vev{(k \cdot a_{(s)})^{n}}.\label{eq:amp1}
\end{align}
We can read out the finite spin coefficients as:
\begin{equation}
	\begin{aligned}
		c_{2m} = \sum_{r=0}^{m} \sum_{j=0}^{m-r} &  \frac{1}{(2 s)!} (-1)^r 2^{2 m - 2 r - 1} (2 s - 2 m)! (-j - \alpha s + s)!\binom {s - s\alpha} {-j + s - s\alpha}^2\binom {-m + r + s}{r} \\
		&r_2^{-j - \alpha s + s}\left(\alpha r_ 1 s + r_ 2 (s - \alpha s) - 1 \right)^{m - r - 1} r_ 1^{j - m + r + \alpha s} \\& \left (-\left (r_ 1 (j - m + r - 2\alpha s) \right) + r_ 2 (j - 2\alpha s + 2 s) - 2 \right) (j - m + r + \alpha s)!\binom {s\alpha} {j - m + r + s\alpha}^2, 
	\end{aligned}
\end{equation}

\begin{equation}
	\begin{aligned}
		c_{2m-1} =& \sum_{r=0}^{m} \sum_{j=0}^{m-r} \frac{1}{(2 s)!} (-1)^r 2^{2 m-2 r-1} \left(j r_2-r_1 (j-m+r)\right) \Gamma (-2 m+2 s+2) (-j-\alpha  s+s)! \\& \binom{s-s \alpha }{-j+s-s \alpha }^2  \binom{-m+r+s}{r}  r_2^{-j-\alpha  s+s} \left(\alpha  \left(r_1-r_2\right) s+r_2 s-1\right)^{m-r-1} r_1^{j-m+r+\alpha  s} \\& (j-m+r+\alpha  s)! \binom{s \alpha }{j-m+r+s \alpha }^2 .
	\end{aligned}
\end{equation}
In the $s \to \infty$ limit, the $r=0$ contribution dominates that gets rid of one summation. Fortunately the other can be evaluated in closed form to give:
\begin{align}
	c_{2m}|_{s \to \infty} &=  \frac{1}{(m!)^2} \alpha ^m r_1^{-m} \left(\alpha  r_1-\alpha  r_2+r_2\right)^m \, _2F_1\left(-m,-m;1;-\frac{(\alpha -1) r_1}{\alpha  r_2}\right),
\end{align} 
\begin{align}
	c_{2m-1}|_{s \to \infty} =& \frac{1}{(m!)^2} \alpha  m r_1^{-m} \left(\alpha  \left(\alpha  r_1-\alpha  r_2+r_2\right)\right){}^{m-1} \left(\left(r_1-r_2\right) \, _2F_1\left(1-m,-m;1;\frac{r_1-\alpha  r_1}{\alpha  r_2}\right) \right.\nn \\
	& \left. +r_2 \, _2F_1\left(-m,-m;1;\frac{r_1-\alpha  r_1}{\alpha  r_2}\right)\right).
\end{align}

\subsection{Exact large $s$ amplitude}
Armed with the infinite $s$ coefficients, we now go back to eq.\eqref{eq:amp1} and perform the summations. The even power contribution evaluates to:
\begin{align}
	\sum_{m=0}^{\infty} c_{2m}|_{s \to \infty} \chi^{2m} = \sum_{m=0}^{\infty} \frac{u^m }{(m!)^2} \, _2F_1(-m,-m;1;z)= I_0(2\sqrt{u}) I_0(2\sqrt{uz}).\label{eq:even}
\end{align}
While the sum with odd terms can be shown to yield:
\begin{align}
		&\sum_{m=1}^{\infty} c_{2m-1}|_{s \to \infty} \chi^{2m-1} \nn \\
		&= \frac{1}{\chi \left(\alpha  r_1-\alpha  r_2+r_2\right)}\bigg[ r_2  \sum_{m=0}^{\infty} \frac{m u^m }{ (m!)^2 } \, _2F_1(-m,-m;1;z) +(r_1-r_2) \sum_{m=0}^{\infty} \frac{m u^m }{ (m!)^2 } \, _2F_1(1-m,-m;1;z) \bigg] \nn \\
	&= \frac{r_2 \sqrt{u\, z}  }{\chi \left(\alpha  r_1-\alpha  r_2+r_2\right)} I_0\left(2 \sqrt{u}\right) I_1\left(2 \sqrt{u z}\right)  + \frac{r_1 \sqrt{u} }{\chi \left(\alpha  r_1-\alpha  r_2+r_2\right)} I_1\left(2 \sqrt{u}\right) I_0\left(2 \sqrt{u \,z}\right). \label{eq:odd}
\end{align}
In the above expressions we have defined:
\begin{align}
	\chi = k \cdot a, \,\, u = \alpha \chi^2 \left( \alpha + \frac{r_2}{r_1} (1-\alpha) \right), \,\, z = \frac{r_1}{r_2} \frac{(1-\alpha)}{\alpha}.
\end{align}
These summation identities are worked out in \S \ref{app:15}. The full amplitude therefore takes the form: 
\begin{align}
{\cal A} &= - \left( \zeta_3 \cdot p_1 \right) \bigg[ I_0(2\sqrt{u}) I_0(2\sqrt{uz}) + \frac{\sqrt{u}}{\chi \left( \alpha \, r_1 - \alpha\, r_2 + r_2\right)} \bigg( r_1 \, I_1( 2\sqrt{u} )  I_0( 2\sqrt{u\,z} ) \nn \\
&+ r_2   \sqrt{z} I_0\left(2 \sqrt{u}\right) I_1\left(2 \sqrt{u z}\right) \bigg)\bigg].\label{eq:amp}
\end{align}
When we take the limit $\alpha \rightarrow 1$ the above reduces to 
\begin{align}
	\mcA^{Regge}  = - (\zeta_3 \cdot p_1) \left( I_0 (2 k \cdot a) + I_1 (2 k \cdot a) \right) \label{eq:regge}
\end{align}
which is consistent with \cite{Cangemi:2022abk}.

\section{String profile}
\label{sec:3}

To describe the classical string, we consider the range of the worldsheet parameter $\sigma \in \left[ 0, \pi a \right]$. Then the charge $g$ is located at the string endpoint $\sigma = \pi a$. We choose the string embedding to be:
\begin{equation}  \label{classical string 1}
	\begin{gathered}
		X^0 = \tau = t,  ~ \vec{X} (t, \sigma) = \left( X^1 (t, \sigma), X^2 (t, \sigma), X^3(t,\sigma) \right) = \frac{1}{2} \left( \vec{F} (t+\sigma) + \vec{F} (t-\sigma) \right) , \\
		\vec{F} (u) = \left( f_1 (u), f_2 (u), f_3(u)  \right), \vec{F}' (u + 2 \pi a) = \vec{F}' (u),
	\end{gathered}
\end{equation}
supplemented with the Virasoro constraints, which imply the following relationships among the embedding functions:
\begin{align}
	&f_1'(t + \pi a) = f_1'(t - \pi  a), ~ f_2'(t + \pi a) = f_2'(t - \pi a),~ f_3'(t + \pi a) = f_3'(t - \pi a),  \label{string constraint 2} \\
	&f_1'(t - \pi  a)^2 + f_2'(t - \pi  a)^2 + f_3'(t - \pi a)^2 = 1. \nonumber
\end{align} 
%
For kinematics we go to the rest frame of the massive string: $p_1 = (E, \mathbf{0})$. We choose the wave vector and the polarization, with the constraints, $k^2 = \varepsilon_k \cdot k = p_1 \cdot k = 0$ to be $k^\mu = E_{\gamma} (0,i,0,i, -\sqrt{2}), ~ \varepsilon_k^\mu = \frac{1}{\sqrt{2}} (1,0,-1,0,0)$.
 The EM current can be expressed as,
\begin{align}  \label{j vector}
	j^\mu (x) = g n^\mu (x) \delta (x - X^1 (t, \pi a)) \delta (y - X^2 (t, \pi a)) \delta (z- X^3 (t, \pi a)),
\end{align}
where, $n^\mu (x) = (\dot{X^0} (t, \pi a), \dot{\vec{X}} (t, \pi a), 0) $. Thus we have,
\begin{equation}  \label{fourier transform 1}
	\begin{aligned}
		\mathcal{FT}\left[\epsilon_k \cdot j\right] &= \int \td^4 x \, e^{i k \cdot x} (\epsilon_k \cdot j (x) ) = \int \td t \, e^{E_\gamma X^1 (t, \pi a)+E_\gamma X^3 (t, \pi a)} \frac{g}{\sqrt{2}} \left( 1 + f_2' (t - \pi a) \right)  \\
		&= \frac{g}{\sqrt{2}} \int_{0}^{2\pi a} \td t \, e^{a\,  E_\gamma f_1(t - \pi  a)+a\,  E_\gamma f_3(t - \pi  a) }  \left( 1 \pm \sqrt{ 1 - f_1'(t - \pi  a)^2-f_3'(t - \pi  a)^2 } \right),
	\end{aligned}
\end{equation}
where in the last line of the above expression, we have used the periodicity of $\vec{F} (u)$ and the expression \eqref{string constraint 2}. Next, we define $\phi = t/a$ and $f_i (t - \pi a) = f_i (a \phi - \pi a) \equiv f_i (\phi)$, for $i=1$ and $3$. Then one obtains:
\begin{align}  \label{FT of f phi expression}
	\frac{\mathcal{FT}\left[\epsilon_k \cdot j\right]}{2 \pi a} = \frac{1}{2 \pi} \frac{g}{\sqrt{2}} \int_{0}^{2 \pi} \td \phi \, e^{a\, E_\gamma \left(f_1(\phi)+f_3(\phi)\right) }  \left( 1 \pm \sqrt{ 1 - \frac{f_1'(\phi)^2}{a^2}- \frac{f_3'(\phi)^2}{a^2} } \right).
\end{align}
It turns out that we need to choose the negative branch of the square root : this is what reproduces the rigid rotating string configuration of the leading Regge trajectory state. For details look at \S Appendix \ref{app:16}.

\subsection{Numerics of the exact profile}
\begin{figure}[h]
	\centering
	\includegraphics[width=1.05 \linewidth]{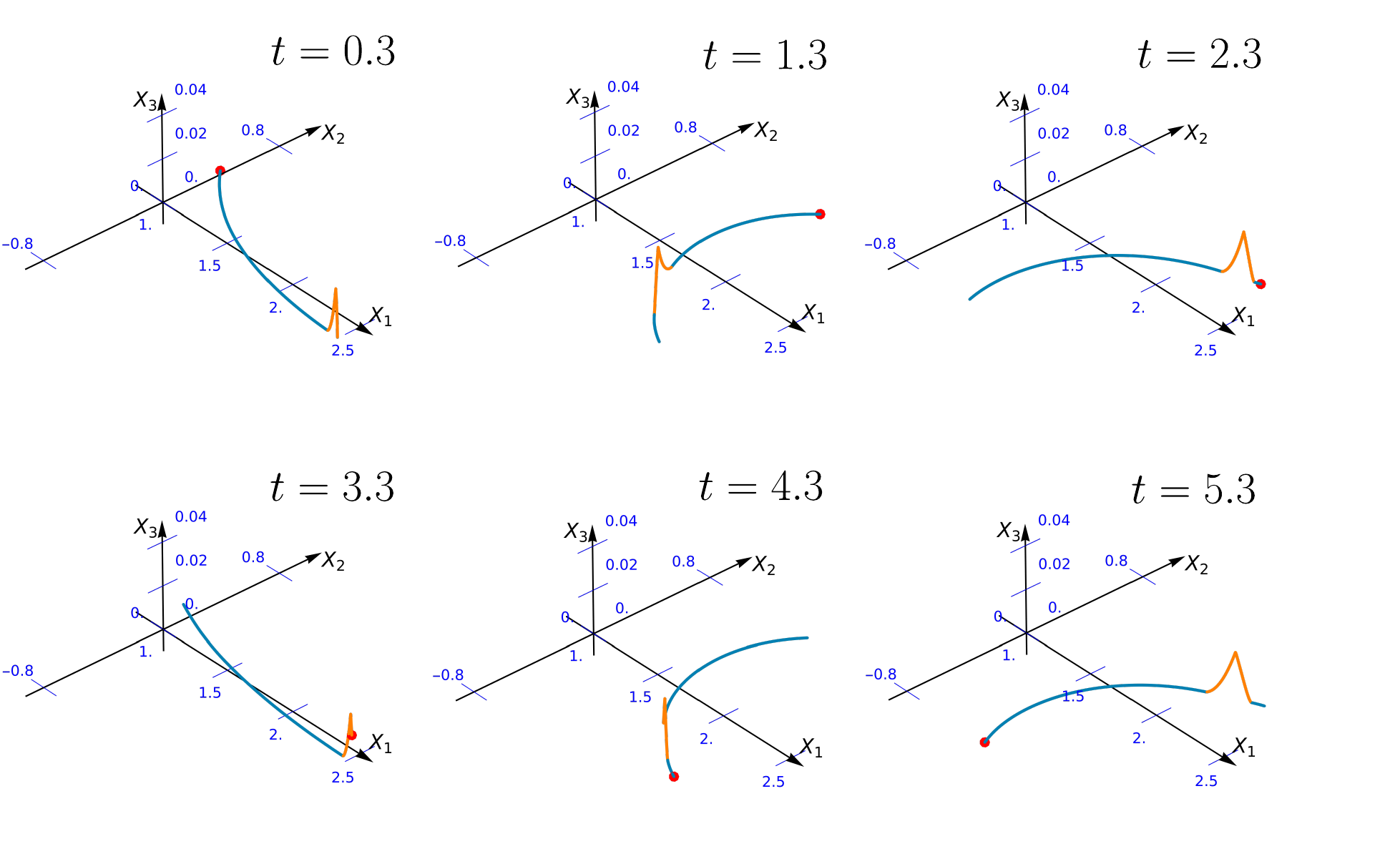}
	\caption{ Snapshots of the time evolution of the classical string profile, for $\alpha = 0.5, r_1 = 1, r_2 = 2$ with $E_{\gamma} =1, a = 1$. The red dot denotes the position of the charge, at $\sigma = \pi a$. The orange parts correspond to the regions $(t \pm \sigma) = 0, 2\pi$, where the string solution extends into the $z$ direction.}
	\label{fig:1}
\end{figure}

Next we look into the classical string profile for the expression \eqref{eq:amp}. Following the strategy for the rigid rotating string, we write the expression as an integral over a variable $\phi$, which gets identified with the one appearing in the r.h.s of eq\eqref{FT of f phi expression}. To this end, we  write one of the Bessel functions using the integral representation, and treat the other Bessel function as the coefficient of the integral. Doing this, we obtain:
\begin{align} \label{lhs-rhs}
		e^{a\,  E_\gamma f_1(\phi)  +a\,  E_\gamma f_3(\phi) }  \left( 1 - \sqrt{ 1 - \frac{f_1'(\phi)^2}{a^2}- \frac{f_3'(\phi)^2}{a^2} } \right) &= e^{-2\sqrt{u}\cos (\phi )} \bigg(-\frac{\sqrt{\alpha  r_1} \cos (\phi ) I_0(2\sqrt{u\, z} )}{\sqrt{\alpha \, r_1 - \alpha \, r_2 + r_2}}\nn \\
		&+\frac{\sqrt{(1-\alpha ) r_2} I_1(2 \sqrt{u \, z })}{\sqrt{\alpha \, r_1 - \alpha \, r_2 + r_2}}+I_0(2 \sqrt{u\, z}  )\bigg).
	\end{align}
We solve this equation numerically. Since we want to get to a string configuration close to $\sqrt{\text{Kerr}}$, we try to confine the string on a two-dimensional plane. While, for the rigid rotating string it was possible to restrict the motion completely on the plane, when higher modes get turned on (as it will happen in our present case) it is inconsistent with the Virasoro constraints to restrict the string completely to the plane \cite{ Alessio:2025nzd}. This shows up in the numerics as an inconsistency if we try to set $f_3(\phi)$ to zero for all $\phi$. See \S appendix \ref{app:2} for further details. This inconsistency will appear at some particular values of the parameter $\phi= \phi_*$ which gets cured by turning on $f_3(\phi_* \pm \epsilon)$ with $\epsilon \rightarrow 0$ locally. This can be done consistently, however while the string is continuous it develops a kink in this transverse $z$ direction : see Fig.\ref{fig:1}.
%
%
%
%
%
%

\begin{figure}[htbp!]
	\centering
	
	\begin{subfigure}[b]{0.45\textwidth}
		\includegraphics[width=0.9\textwidth]{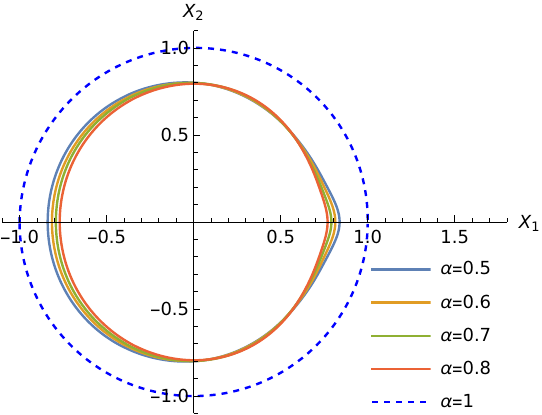}
		\caption{for different values of $\alpha = 0.5, 0.6, 0.7, 0.8, 1$, with $r_1 = 1, r_2 = 2$.}
		\label{fig: charge trajectory different alpha}
	\end{subfigure}
	\hfill
	\begin{subfigure}[b]{0.45\textwidth}
		\includegraphics[width=0.9\textwidth]{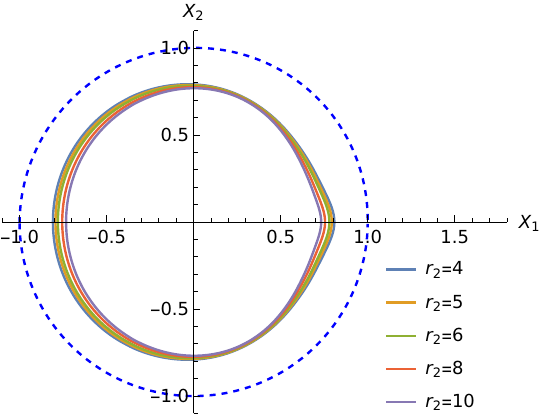}
		\caption{for different values of $r_2 = 4,5,6,8,10$, with $r_1 = 1, \alpha = 0.5$. The dashed circle denotes the trajectory for the Regge case ($\alpha = 1$).}
		\label{fig: charge trajectory different r1r2}
	\end{subfigure}
		\caption{Projection of the trajectory of the point charge on the $X_1- X_2$ plane for $E_{\gamma} =1, a = 1$. }
	\label{fig: charge trajectory full}
\end{figure}

In Fig.\ref{fig: charge trajectory different alpha} we plot the two dimensional projection of the trajectory of the charge which is attached to the string, for different values of $\alpha$. As expected $\alpha=1$ traces out the circle corresponding to the rigid rotating string. As $\alpha \rightarrow 1/2$ we find deformations of the trajectory which now starts to explore more area of the disk. In Fig.\ref{fig: charge trajectory different r1r2}  we look at the trajectory by fixing $r_1 =1$ and $\alpha = 1/2$ while varying $r_2$. As the state deviates from zero-depth, i.e., $r_2$ increases we find the trajectory explores more of the interior region of the disk. 
%

\newpage

\section{Conclusions}
\label{sec:4}
Highly excited string states are expected to describe black hole microstates. This expectation has been gathering evidence at least in the correspondence point. In the open string context the analog of spinning black holes is the $\sqrt{\text{Kerr}}$ state which has a well-defined three point amplitude with the photon. Motivated by this amplitude we investigated the classical charge configuration corresponding to the three point amplitude $\vev{ \Phi^s_1 \Phi^s_2 \gamma}$.  For the excited physical states in bosonic string theory we considered the DDF operators of the form $\Phi^s \sim A_{-r_1}^{s_1} A_{-r_2}^{s_2}$ and evaluated the amplitude for arbitrary integers $r_i, \, s_i$. We have chosen the polarization tensors such that the DDF state corresponds to a superposition of spinning massive states in the spacetime whose minimal spin is bounded by the total length of partitions, $s = s_1 + s_2$, appearing in $\ket{\Phi^s}$. Thereafter we extracted the finite $s_i$ ``spin'' coefficients in  closed form and took the $s \rightarrow \infty$ limit to match classical observables. 
\begin{figure}[h]
	\centering
	\includegraphics[width= .45\linewidth]{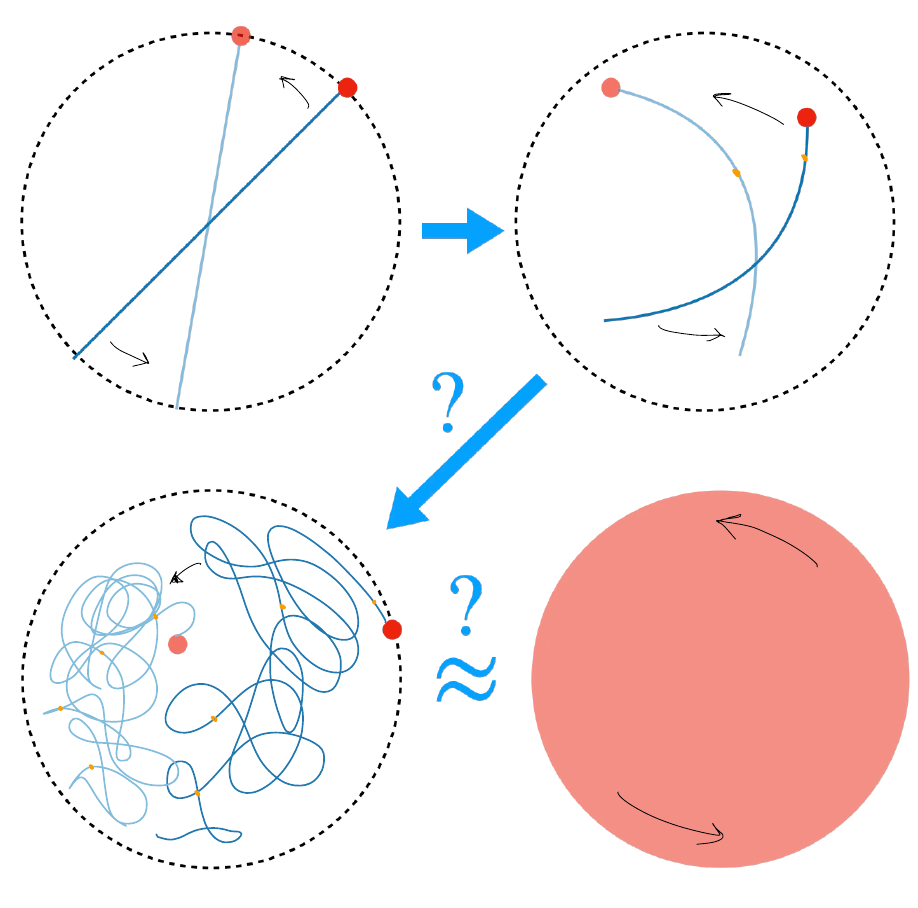}
	\caption{A cartoon of the classical string profile. The charge is indicated with the red endpoint. The orange dots indicate location of the kink. The slightly transparent afterimage indicates the string-charge configuration in a future instant. The lower left indicates the $\ket{\sqrt{\text{Kerr}}}$ configuration of charged rotating disk.}
	\label{fig:cartoon}
\end{figure}

The  configuration corresponding to $\ket{\sqrt{\text{Kerr}}}$ is a uniformly charged rotating disk \cite{Arkani-Hamed:2019ymq} (lower-right Fig.\ref{fig:cartoon}) where as the excited string state on the leading Regge trajectory in the classical limit generates the same multipole expansion as that of a rigid rotating string with charge at one endpoint (upper-left Fig.\ref{fig:cartoon}). Moving away from the leading Regge trajectory, we find that the rotating string is no longer rigid. A consequence of trying to confine the string to a two dimensional plane is the development of a moving kink at a point which is directed in the transverse direction. This is indicated in upper-right of Fig.\ref{fig:cartoon}.  The motion of the charge particle now starts to explore inside of the circle which the leading Regge configuration had traced out. It is thus conceivable that as we keep exciting more DDF oscillators, the string endpoint will explore a greater area of the disk while becoming floppy at the same time. In this manner it may start approximating the configuration of $\ket{\sqrt{\text{Kerr}}}$. It will of course be interesting to find the approach to this configuration in case it exists within the realm of string theory. 

\subsection*{Acknowledgements}
D.D. is thankful to members of the CHEP, IISc-Bangalore who were in audience when parts of this result was presented, for their feedback and questions.

\appendix
\renewcommand{\theequation}{\thesection.\arabic{equation}}

\section{Details of $\vev{\Phi^s_1 \Phi^s_2 \gamma}$ computation}
\label{app:1}

In section \S\ref{sec:2} the vertex operators are showed in terms of simplified notations:
\begin{align}
	a_1(r_1,m_1) &= \frac{-i \zeta_1 \cdot \partial^{m_1} X_1}{(m_1-1)!} ; \nonumber \\
	b_1(r_1,m_1) &= S_{r_1-m_1}(-\frac{r_1 i}{s_1!} q_1 \cdot \partial^{s_1} X_1); \nonumber \\
	a_2(r_2,m_2) &= \frac{i \zeta_2 \cdot \partial^{m_2} X_2}{(m_2-1)!} ; \nonumber \\
	b_2(r_2,m_2) &= S_{r_2-m_2}(-\frac{r_2 i}{s_2!} q_2 \cdot \partial^{s_2} X_2); \nonumber \\
	a_3(1,1)&= i \zeta_3 \cdot \partial X_3; \nonumber \\
	c_j &= e^{i~p_j \cdot X_j}.
\end{align}
Here $\zeta_i$-s are related to the DDF polarization vectors $\lambda_i$s in following way,
\begin{align}
	\zeta_i= \lambda_i-(\lambda_i \cdot p_i) q_i.
\end{align}
\noindent Each of the DDF operators $A_{-r}$ acting on the $i$-th vacuum state have DDF photon momentum $-r~q_i$. For the $\ket{\Phi^s_i}$ states considered in \eqref{HES-states} all the DDF photon momenta are parallel i.e. $q_1=-q_2=q_3=q$ with $q^2=0$. This implies that: 
\begin{align}
\zeta_i \cdot q_j &= 0. \label{eq:cons}
\end{align}

\subsection{OPE of DDFs}

The only non-zero contractions between the several constituent elements of the vertex operators are listed down below. We have simplified our notation further using the fact that the $b_i(r,n_i)$-s contain only the Sch\"{u}r polynomials having $-\frac{i~r~ q \cdot \partial^s X_i}{s!}$ as arguments. This part can only be contracted with the $e^{i~ p_j \cdot X_j}$ part of other vertex operators ($i \neq j$) for our restricted kinematic choice, eq.\eqref{eq:cons}. These contractions of $b_i$'s result only in numerical factors which we have directly written down in what follows. Ultimately, the full contraction simplifies into the following six different possible complete contractions:
\begin{align}
	&\sum_{m_1,m_2=1}^{r_1,r_2} \left( a_1(r_1,m_1)a_2(r_2,m_2)\right) \left(b_1(r_1,m_1) c_{2,3} \right) \left( b_2(r_2,m_2) c_{1,3} \right) \bigg \rvert_{z \in \{0,1, \infty \}} \nonumber \\
	&= \zeta_1 \cdot \zeta_2 (-1)^{r_1} r_1 \delta_{r_1,r_2} = \left( a_1(r_1)a_2(r_2) \right). \nonumber \\
	&\sum_{m_1=1}^{r_1} \left( a_1(r_1,m_1)c_{2,3}\right) \left(b_1(r_1,m_1) c_{2,3} \right) \bigg \rvert_{z \in \{0,1, \infty \}}= (a_1(r_1)c))=(-1)^{r_1+1} \zeta_1 \cdot p_2. \nonumber \\
	&\sum_{m_2=1}^{r_2} \left( a_2(r_2,m_2)c_{1,3}\right) \left(b_2(r_2,m_2) c_{1,3} \right) \bigg \rvert_{z \in \{0,1, \infty \}}= (a_2(r_2)c))=-\zeta_2 \cdot p_1. \nonumber \\
	&|z_{12}|^{2N-1} |z_{13}||z_{23}| \sum_{m_1=1}^{r_1} \left( a_1(r_1,m_1)a_3\right) \left(b_1(r_1,m_1) c_{2,3} \right) \nonumber \\
	&~~~~~~~~~~= |z_{12}|^{2N-1} |z_{13}||z_{23}| (a_1(r_1)a_3) = (-1)^{r_1} r_1 \zeta_1 \cdot \zeta_3 . \nonumber \\
	&|z_{12}|^{2N-1} |z_{13}||z_{23}| \sum_{m_2=1}^{r_2} \left( a_2(r_2,m_2)a_3\right) \left(b_2(r_2,m_2) c_{1,3} \right) \nonumber \\
	&~~~~~~~~~~=|z_{12}|^{2N-1} |z_{13}||z_{23}| (a_2(r_2)a_3) = - r_2 \zeta_2 \cdot \zeta_3.\nonumber \\
	&|z_{12}|^{2N-1} |z_{13}||z_{23}| (a_3(1,1)c_{1,2}) = \zeta_3 \cdot p_1. \label{eq:ope}
\end{align}
In the above each contraction is showed inside a parenthesis. For example the contraction $\langle -i \zeta_1 \cdot \partial X_1 ~ i \zeta_2 \cdot X_2 \rangle = (a_1(1,1) a_2(1,1))=(a_1(1)a_2(1))$ using our compact notation. Similarly another contraction  $\langle -i \zeta_1 \cdot \partial X_1 ~ (e^{i~p_1 \cdot X_1}+e^{i~p_2 \cdot X_2}) \rangle = (a_1(1,1) c_{2,3})=(a_1(1)c)$
\subsection{Amplitude Calculation}
Using the above notations the amplitude \eqref{amphhp} can be expressed as: 
\begin{align}
	&\langle:\big( a_1(r_1) \big)^{N_1}\big( a_1(r_2) \big)^{N_2} c_1: ~ :\big( a_2(r_1) \big)^{N_1}\big( a_2(r_2) \big)^{N_2} c_1: ~ :a_3 c_3: \rangle \nonumber \\
	&= \bigg[ \sum_{k,l=1}^{N_1,N_2} \binom{N_1}{k}^2\binom{N_2}{l}^2~k! ~l!~ \big( a_1(r_1)a_2(r_1)\big)^k \big( a_1(r_2)a_2(r_2)\big)^l\Bigg( \big(a_1(r_1)c\big)~\big(a_1(r_1)c\big) \Bigg)^{N_1-1-k} \nonumber \\
	&~~~~~~~~~~\Bigg( \big(a_1(r_2)c\big)~\big(a_2(r_2)c\big) \Bigg)^{N_2-1-l} \bigg] \times \nonumber \\
	&\Bigg[\big(a_3c \big) \Bigg( \big(a_1(r_1)c\big)~\big(a_1(r_1)c\big) \Bigg)\Bigg( \big(a_1(r_2)c\big)~\big(a_2(r_2)c\big) \Bigg) \nonumber \\
	&~~~~+(N_1-k) \Bigg(\big(a_1(r_1)a_3\big)~\big(a_2(r_1)c\big)+\big(a_2(r_1)a_3\big)~\big(a_1(r_1)c\big) \Bigg)\Bigg( \big(a_1(r_2)c\big)~\big(a_2(r_2)c\big) \Bigg) \nonumber \\
	&~~~~+ (N_2-l)\Bigg(\big(a_1(r_2)a_3\big)~\big(a_2(r_2)c\big)+\big(a_2(r_2)a_3\big)~\big(a_1(r_2)c\big) \Bigg) \Bigg( \big(a_1(r_1)c\big)~\big(a_1(r_1)c\big) \Bigg) \Bigg] \times \prod_{i<j} |z_{ij}|^{p_i \cdot p_j}. \label{eq:ampN}\\
	&= \text{amp} (N_1,N_2,r_1,r_2)   \prod_{i<j} |z_{ij}|^{p_i \cdot p_j}. \nn
\end{align}
In the above expression the first term of the second square bracket corresponds to the case where $a_3$ is contracted with the operators $c$, and out of the remaining $N_1$ number of $a_1(r_1)$ and $a_2(r_1)$-s $k$ number of $a_1(r_1)$ are contracted with $k$ number of $a_2(r_1)$. The rest are contracted with the $c$-s. This can produce $\binom{N_1}{k}^2 k!$ identical terms. Similar combinatorial argument also holds for $r_1 \leftrightarrow r_2$ interchange. 

The other terms are similarly obtained via similar combinatorics. In the $2$nd term the $a_3$ is contracted with either of the $a_i(r_1)$-s and in the $3$rd term $a_3$ contracted with either of the $a_i(r_2)$-s. Since $a_i(r_1)$ cannot be contracted with $a_j(r_2)$ when $r_1 \neq r_2$, these are the only possible contractions for our kinematic choice. 

Next we need to perform the $\text{d}z_1 \text{d}z_2 \text{d}z_3$ integral. Using the momentum on-shell relation and momentum conservation in the total amplitude,
\begin{align}
    &\prod_{i<j} |z_{ij}|^{p_i \cdot p_j} = |z_{12}|^{2 (N-1)} \text{, where } N=N_1r_1+N_2r_2 \nonumber \\
    &\text{d}z_1 \text{d}z_2 \text{d}z_3~ \prod_{i<j} |z_{ij}|^{p_i \cdot p_j} = \frac{\text{d}z_1 \text{d}z_2 \text{d}z_3}{|z_{12}| |z_{13}| |z_{23}|} ~ |z_{12}|^{2N-1} |z_{13}||z_{23}|.
\end{align}
In the above expression $\frac{\text{d}z_1 \text{d}z_2 \text{d}z_3}{|z_{12}| |z_{13}| |z_{23}|}$ is a $SL(2,R)$ invariant factor which can be taken care of by fixing the three $z_i$ coordinates using the symmetry. We chose $z_1=0, ~z_2=1$ and $z_3=\infty$, this choice is denoted as $z\in \{ 0,1,\infty\}$ throughout (including the evaluations in eq.\eqref{eq:ope}. With these choice the amplitude \eqref{amphhp} becomes
\begin{align}
    \mathcal{A}&= \int \text{d}z_1 \text{d}z_2 \text{d}z_3 \langle:\big( a_1(r_1) \big)^{N_1}\big( a_1(r_2) \big)^{N_2} c_1: ~ :\big( a_2(r_1) \big)^{N_1}\big( a_2(r_2) \big)^{N_2} c_1: ~ :a_3 c_3: \rangle \nonumber \\
    &= |z_{12}|^{2N-1} |z_{13}||z_{23}| \times \text{amp} (N_1,N_2,r_1,r_2) \bigg \rvert_{z \in \{ 0,1,\infty\}}
\end{align}
where $\text{amp} (N_1, N_2, r_1, r_2)$ has been defined in the last equality of eq.\eqref{eq:ampN}. Once we combine and take the $(0,1,\infty)$ limits, we get the r.h.s of eq.\eqref{amphhp}.

\section{Summation identities}
\label{app:15}
In this section we give a derivation of the identities eq.\eqref{eq:even} and eq.\eqref{eq:odd}. 
\subsubsection*{Calculation of the even sum}
\begin{figure}
	\centering
	\includegraphics[width=\textwidth]{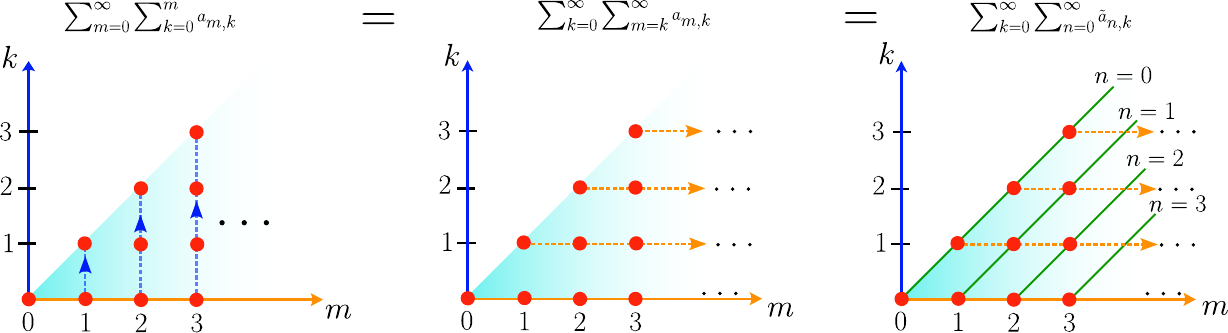}
	\caption{Details of the reordering of the summation indices. The red dots denote the $(m,k)$ indices of the summation. From the first to the second diagram, one counts along the $m$ axis instead of the $k$ axis. In the last figure, one counts the different values of $n = m-k$ for a fixed $k$. }
	\label{fig: string sum}
\end{figure}

First we want to prove: 
\begin{align}
S_e(u,z) &\equiv \sum_{m=0}^{\infty} \frac{u^m }{(m!)^2} \, _2F_1(-m,-m;1;z)= I_0(2\sqrt{u}) I_0(2\sqrt{uz}) .
 \end{align}
Using series representation of hypergeometric function : $_2F_1(-m,-m;1;z) = \sum_{k=0}^m \binom{m}{k}^2 z^k$ we can express
\begin{align}
	S_e (u,z) &= \sum_{m=0}^\infty \frac{u^m}{(m!)^2} \sum_{k=0}^m \binom{m}{k}^2 z^k .
\end{align}
Next, we interchange the order of the two summations. Then we have:
\begin{equation}
	\begin{split}
		S_e(u,z) &= \sum_{k=0}^\infty  \sum_{m=k}^\infty \binom{m}{k}^2 z^k \frac{u^m}{(m!)^2} .
	\end{split}
\end{equation}
A further change of variables to $n= m-k$ decouples the sums: 
\begin{equation}
	\begin{split}
		S_e(u,z) &= \sum_{k=0}^\infty \frac{(uz)^k}{(k!)^2} \sum_{n=0}^\infty \frac{u^n}{(n!)^2}, 
\end{split}
\end{equation}
The details of the rearrangement of the indices in the above expression is outlined in Fig.\ref{fig: string sum}. Now using the series expression for the modified Bessel function:
\begin{equation}
	I_0(2\sqrt{\chi}) = \sum_{n=0}^\infty \frac{\chi^n}{(n!)^2},
\end{equation}
we obtain,
\begin{align}
	S_e(u,z) = I_0(2\sqrt{u}) \cdot I_0(2\sqrt{uz}).
\end{align}

\subsubsection*{Calculation of the odd sum}
The computation of the odd sums, eq.\eqref{eq:odd} is relatively easy as they follow from the even sum.
First we evaluate:
\begin{equation}
	\begin{split}
		S^{1}_{o} (u, z) &\equiv \sum_{m=0}^{\infty} \frac{m u^m }{ (m!)^2 } \, _2F_1(-m,-m;1;z)  = u \dfrac{d}{d u} \left(S_e(u,z)\right) \\
		&= \sqrt{u} \left(I_1\left(2 \sqrt{u}\right) I_0\left(2 \sqrt{u z}\right)+\sqrt{z} I_0\left(2 \sqrt{u}\right) I_1\left(2 \sqrt{u z}\right) \right).
	\end{split} 
\end{equation}

\noindent The other sum which appears in the odd part is  
\begin{align}
S^{2}_{o} (u, z)&\equiv \sum_{m=0}^{\infty} \frac{m u^m }{ (m!)^2 } \, _2F_1(1-m,-m;1;z).
\end{align}
Once again we use the series representation of ${}_2F_1$:
\begin{align}
	_2F_1(1-m,-m;1;z) = \sum_{k=0}^{m-1} \binom{m}{k} \binom{m-1}{k}  z^k.
\end{align}

\noindent Then,
\begin{equation}
	\begin{split}
		S^{2}_{o} (u, z) &= \sum_{m=0}^{\infty} \frac{m u^m }{ (m!)^2 } \, _2F_1(1-m,-m;1;z) 
		= \sum_{k=0}^{\infty} \frac{(u z)^k}{(k!)^2} \sum_{n=0}^{\infty} \frac{n u^n}{(n!)^2} \\
		&= I_0\left(2 \sqrt{u z}\right) \cdot u \dfrac{d}{d u} \left( I_0\left(2 \sqrt{u}\right) \right) \\
		&= \sqrt{u} \, I_1\left(2 \sqrt{u}\right) I_0\left(2 \sqrt{u z}\right).
	\end{split}
\end{equation}
These proves the identities used in \S\ref{sec:2}. 

\section{Regge state and the rigid rotating string}
\label{app:16}
We want the rigid rotating string profile to reproduce eq\eqref{eq:regge}.
The rigid rotating string is given by the embeddings,
\begin{align}
	X(\tau, \sigma) = \left(\tau, a \cos\frac{\tau}{a} \cos\frac{\sigma}{a}, a \sin\frac{\tau}{a} \cos\frac{\sigma}{a}, 0 \right),
\end{align}
and hence:
\begin{align*}
	n^\mu (\tau / a) = \dot{X}^\mu (\tau, \pi a) = \left(1, \sin\frac{\tau}{a} , -\cos\frac{\tau}{a} , 0 \right).
\end{align*}
 Then following \eqref{j vector}, the current 4-vector is:
\begin{align}
	j^\mu (x) = g n^\mu (x) \delta (x + a \cos(t/a)) \delta (y + a \sin(t/a)) \delta (z).
\end{align}
The last equality is obtained by considering a charge $g$ introduced at one of the endpoints of a rigid rotating string. Then the Fourier transform of $\varepsilon_k \cdot j$ is:
\begin{equation} \label{FT of rigid rotating 1}
	\begin{split}
		\frac{\mathcal{FT}\left[\epsilon_k \cdot j\right]}{2 \pi a} &= \frac{1}{2 \pi} \frac{g}{\sqrt{2}} \int_{0}^{2\pi} \td \phi \, e^{- a E_\gamma \cos(\phi)} \left( 1 - \cos (\phi) \right) = \frac{g}{\sqrt{2}} \left( I_0 (a E_{\gamma}) + I_1 (a E_{\gamma}) \right),
	\end{split}
\end{equation}
where, we have substituted $\phi = t/a$ as earlier. Now, for the rigid rotating string solution, one has,
\begin{align}
	f (\phi) = f_1 (t - \pi a) = X^1 (t, \pi a) = - a \cos (t/a) = -a \cos (\phi).
\end{align}
\noindent Substituting this in \eqref{FT of f phi expression}, we obtain:
\begin{align}
	\frac{\mathcal{FT}\left[\epsilon_k \cdot j\right]}{2 \pi a} = \frac{1}{2 \pi} \frac{g}{\sqrt{2}} \int_{0}^{2\pi} \td \phi \, e^{- a E_\gamma \cos(\phi)} \left( 1 \pm \cos (\phi) \right).
\end{align}
By comparing the above with the expression \eqref{FT of rigid rotating 1}, we see that one needs to take the negative branch in the expression \eqref{FT of f phi expression}. Thus one finally obtains,
\begin{align}  \label{FT of f phi expression final}
	\frac{\mathcal{FT}\left[\epsilon_k \cdot j\right]}{2 \pi a} = \frac{1}{2 \pi} \frac{g}{\sqrt{2}} \int_{0}^{2 \pi} \td \phi \, e^{E_\gamma f_1(\phi) }  \left( 1 - \sqrt{ 1 - \frac{f_1'(\phi)^2}{a^2} } \right).
\end{align}

\section{Numerical details}  \label{app:2}

\noindent Here we present the numerical details for the solution $f_1(\phi)$ and  $f_3(\phi)$ which dictate the string profile and follow eq.\eqref{lhs-rhs}. As we have already mentioned, we want to restrict the string onto the two dimensional plane as much as possible in order to be closer to the $\sqrt{\text{Kerr}}$ configuration. Therefore firstly we turn off $f_3(\phi)$ and figure out the regions where having only $f_1(\phi)$ in eq\eqref{lhs-rhs} will  incur inconsistencies. Thus we attempt to numerically solve:
\begin{align} \label{to solve numerical}
	e^{a\,  E_\gamma f_1(\phi) }  \left( 1 - \sqrt{ 1 - \frac{f_1'(\phi)^2}{a^2}} \right) &= e^{-2\sqrt{u}\cos (\phi )} \bigg(-\frac{\sqrt{\alpha  r_1} \cos (\phi ) I_0(2\sqrt{u\, z} )}{\sqrt{\alpha \, r_1 - \alpha \, r_2 + r_2}}\nn \\
	&+\frac{\sqrt{(1-\alpha ) r_2} I_1(2 \sqrt{u \, z })}{\sqrt{\alpha \, r_1 - \alpha \, r_2 + r_2}}+I_0(2 \sqrt{u\, z}  )\bigg).
\end{align}
Without the loss of generality, we fix $a = 1, E_\gamma = 1$. Since $f_1(\phi)$ has a $2\pi$ periodicity, we expand $f_1(\phi)$ as a Fourier series:
\begin{align}  \label{fourier series}
	f_1(\phi) = \mathsf{a}_0 + \sum_{n=0}^{n_\text{max}}  \mathsf{b}_n \sin (n \phi) + \mathsf{a}_n \cos (n \phi),
\end{align}
with the mode cut-off $n_\text{max}$ finite. We then try to minimize the mean RMS value of the difference between the l.h.s. and the r.h.s. (called the `residual') in \eqref{to solve numerical} over the range $\phi \in \left[0, 2\pi\right]$, by varying the Fourier coefficients $\mathsf{a}_n, \mathsf{b}_n$. Thus the approximate numerical solution for a fixed $n_{\text{max}}$ is obtained. To find the most optimized solution, we then scan over a range of $n_{\text{max}}$ (we have scanned till $n_{\text{max}} = 20$), for which the RMS residual is the smallest. This provides us with the final optimized solution of $f_1(\phi)$.

\begin{figure}[htbp!]
	\centering
	\includegraphics[width=0.4\textwidth]{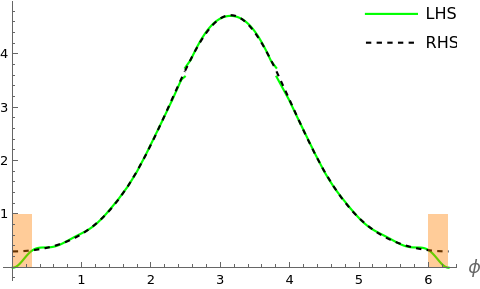}
	\caption{l.h.s. and r.h.s. of eq\eqref{lhs-rhs} with numerical solutions of $f_1(\phi)$ for $\alpha = 0.5, r_1 = 1, r_2 = 2$. Without introducing non-zero $f_3(\phi)$ there is a mismatch at $\phi=0$ and $\phi=2 \pi$. }
	\label{fig: lhs-rhs}
\end{figure}

Note that the r.h.s. of \eqref{to solve numerical} is invariant under $\phi \to (2\pi - \phi)$. This implies $f_1(\phi)$ is symmetric about $\phi = \pi$, which subsequently implies that the $\mathsf{b}_n$ coefficients exactly zero in \eqref{fourier series}. Indeed, in the obtained numerical solutions the $\mathsf{b}_n$ coefficients turn out infinitesimal compared to the $\mathsf{a}_n$ coefficients. In Fig.\ref{fig: lhs-rhs}, we plot the l.h.s. and the r.h.s. of \eqref{to solve numerical} for $\alpha = 0.5, r_1 = 1, r_2 = 2$. Note, around $\phi = \phi_* = 0, 2\pi$ the solution becomes inconsistent. Thus, around these $\phi_*$ values, one has to solve \eqref{lhs-rhs} by including $f_3(\phi)$ and its derivative, to obtain a consistent solution. To do this, we use the solution of $f_1 (\phi)$ obtained from \eqref{to solve numerical}, and replace it in \eqref{lhs-rhs}.

\begin{figure}[h]
	\centering
	\begin{subfigure}[b]{0.4\textwidth}
	\includegraphics[width=\textwidth]{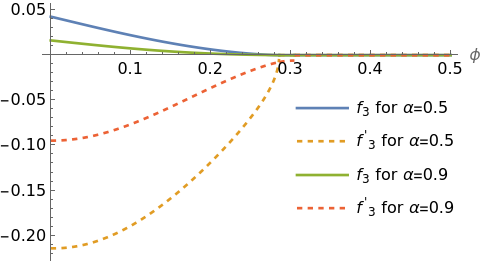}
	\caption{Numerical solution of $f_3(\phi)$ and its derivative near $\phi = 0$ for different values of $\alpha$. We have chosen $r_1=1$ and $r_2=2$.  }
	\label{fig:f3a}
	\end{subfigure}
	\hfill
	\begin{subfigure}[b]{0.4\textwidth}
	\includegraphics[width=\textwidth]{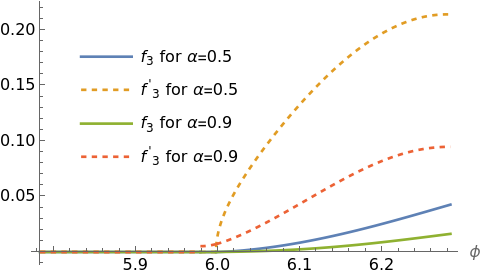}
	\caption{Numerical solution of $f_3(\phi)$ and its derivative near $\phi = 2\pi$ for different values of $\alpha$.}
	\label{fig:f3b}
	\end{subfigure}

	\caption{The numerical solution to $f_3(\phi)$. }
	\label{fig:f3plot}
\end{figure}

The numerically obtained profile of $f_3(\phi)$ and its derivative are plotted in Fig.\ref{fig:f3plot}. As expected, we notice that as we are closer to the Regge state ($\alpha=1$) the extent of this portion of the string in the third direction diminishes. 

\begin{figure}[h]
	\centering
	\begin{subfigure}[b]{0.45\textwidth}
	\includegraphics[width=\textwidth]{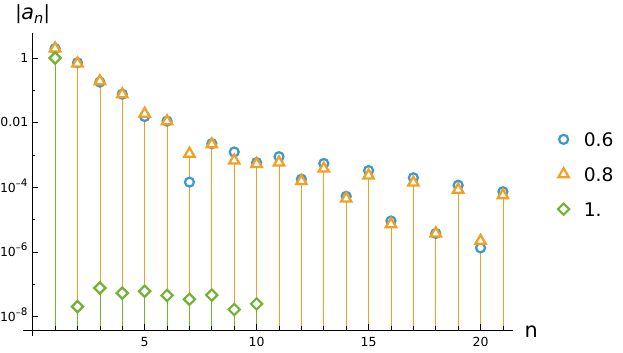}
	\caption{Plot of $|\mathsf{a}_n|$ vs $n$ for $\alpha = 0.6, 0.8, 1$. Here, $r_1 = 1, r_2 = 2$ with $E_{\gamma} =1, a = 1$. The $\mathsf{b}_n$ coefficients are omitted since those are exponentially small (maximum $\sim \mathcal{O} (10^{-7})$) compared to the $\mathsf{a}_n$ coefficients.}
	\label{fig: fourier coeffs}
	\end{subfigure}
	\hfill
	\begin{subfigure}[b]{0.45\textwidth}
	\includegraphics[width=\textwidth]{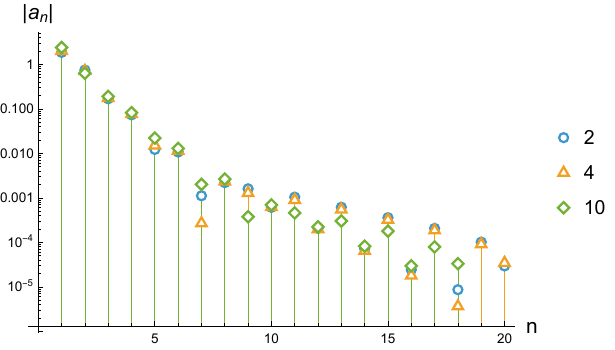}
	\caption{Plot of $|\mathsf{a}_n|$ vs $n$ for $r_2 = 2, 4, 10$ with $r_1 = 1, \alpha = 0.5, E_{\gamma} =1, a = 1$.}
	\label{fig: fourier coeffs alpha}
	\end{subfigure}

	\caption{Fourier coefficients of $f_1(\phi)$. }
	\label{fig: numerics}
\end{figure}

In Fig.\ref{fig: fourier coeffs}, we plot the Fourier coefficients $\mathsf{a}_n$ as a function of $n$ for different values of $\alpha$ with $r_1 = 1, r_2 = 2$. For the Regge state, the coefficient $\mathsf{a}_1$ is order 1, while the rest of the coefficients are order $10^{-7} $(note that for the analytical solution the rest of the coefficients are exactly zero). On the other hand, for the non-Regge states, there is a clear mixing between different modes, instead of one dominant coefficient in the Fourier series. In Fig.\ref{fig: fourier coeffs alpha}, we plot the Fourier coefficients $\mathsf{a}_n$ as a function of $n$ for $r_2 = 2, 4, 10$ with $r_1 = 1, \alpha = 0.5, E_{\gamma} =1, a = 1$. The Fourier coefficients do not show much variation with the change of $r_2$, signifying that the string profile remains similarly floppy throughout this class of states.

\bibliographystyle{jhep}
\bibliography{references}

\end{document}